\documentclass[reprint,prl,twocolumn,showpacs,superscriptaddress,aps,longbibliography]{revtex4-1}

\usepackage[utf8]{inputenc}

\usepackage[colorlinks=true,citecolor=blue]{hyperref} 
\usepackage{graphicx}
\usepackage{bm}
\usepackage{amsmath}
\usepackage{amssymb}

\usepackage{graphicx}
\DeclareGraphicsExtensions{.png,.jpg,.eps}

\usepackage{float}
\usepackage{xcolor}
\usepackage{amsmath}
\usepackage[printwatermark]{xwatermark}
\usepackage{float}
\usepackage[colorlinks=true,citecolor=blue]{hyperref}
\DeclareMathAlphabet\mathbfcal{OMS}{cmsy}{b}{n}

\usepackage{xcolor}

\begin{document}

\title{Anharmonicity reveals the tunability of the charge density wave orders in monolayer VSe$_2$} 

\author{Adolfo O. Fumega}
\email{adolfo.oterofumega@aalto.fi}
\affiliation{Department of Applied Physics, Aalto University, 02150 Espoo, Finland}

\author{Josu Diego}
\affiliation{Fisika Aplikatua Saila, Gipuzkoako Ingeniaritza Eskola, University of the Basque Country (UPV/EHU), San Sebastián, Spain}
\affiliation{Centro de Física de Materiales (CSIC-UPV/EHU), San Sebastián, Spain}

\author{V. Pardo}
\affiliation{Departamento de Física Aplicada, Universidade de Santiago de Compostela, Santiago de Compostela, Spain}
\affiliation{Instituto de Materiais iMATUS, Universidade de Santiago de Compostela, Santiago de Compostela, Spain}

\author{S. Blanco-Canosa}
\affiliation{Donostia International Physics Center (DIPC), San Sebastián, Spain}
\affiliation{IKERBASQUE, Basque Foundation for Science, 48013 Bilbao, Spain}

\author{Ion Errea}
\email{ion.errea@ehu.eus}
\affiliation{Fisika Aplikatua Saila, Gipuzkoako Ingeniaritza Eskola, University of the Basque Country (UPV/EHU), San Sebastián, Spain}
\affiliation{Centro de Física de Materiales (CSIC-UPV/EHU), San Sebastián, Spain}
\affiliation{Donostia International Physics Center (DIPC), San Sebastián, Spain}

\begin{abstract}

VSe$_2$ is a layered compound that has attracted great attention due to its proximity to a ferromagnetic state that is quenched by the presence of a charge density wave (CDW) phase. In the monolayer limit, unrelated experiments have reported different CDW orders with transition temperatures in the range of 130 to 220 K, making this monolayer very controversial.
Here we perform first principles non-perturbative anharmonic phonon calculations in monolayer VSe$_2$ in order to estimate the CDW order and the corresponding transition temperature.
Our analysis solves previous experimental contradictions as we reveal that monolayer VSe$_2$ develops two independent charge density wave orders associated to $\sqrt{3} \times \sqrt{7}$ and $4 \times 4$ modulations that compete as a function of strain. 
In fact, tiny variations of only 1.5\% in the lattice parameter are enough to stabilize one order or the other, which underlines that the CDW order becomes substrate-dependent. 
The predicted CDW temperature is strain-dependent and has a value of around 220 K, in good agreement with experiments. 
Moreover, we analyze the impact of external Lennard-Jones interactions on the CDW. We show that these can act together with the anharmonicity to suppress the CDW orders. In the particular case of monolayer VSe$_2$, this may give rise to the emergence of a ferromagnetic order. 

\end{abstract}


\maketitle

\section{Introduction}
Two-dimensional (2D) materials are an ideal platform to artificially engineer heterostructures with new functionalities due to the weak van der Waals bonding between layers \cite{vdwHT2013}. 
Monolayers hosting symmetry-broken phases, such as superconductivity \cite{NbSe22015,vano2021evidence}, magnetism \cite{doi:10.1021/acs.nanolett.6b03052,Huang2017,Gong2017,Fei2018,doi:10.1021/acs.nanolett.9b00553}, ferroelectricity \cite{doi:10.1021/acs.nanolett.7b04852, Yuan2019}, charge density waves (CDWs) \cite{PhysRevB.100.241404,Chen2017}, or multiferroicity \cite{Song2022, Fumega_2022}, represent the most interesting building blocks to design novel phases of matter. 
One of the main challenges in the task of engineering novel functional materials with broken-symmetry monolayers is to overcome the restrictions imposed by the reduced dimensionality \cite{PhysRev.158.383,PhysRevLett.17.1133}, which may prevent the formation of these phases, and the competition between ordered phases due to the subtle interplay of different interactions \cite{Du2021,Li2021}.
For instance, CDW phases have been reported to destroy \cite{doi:10.1021/acs.jpcc.9b08868,doi:10.1021/acs.jpcc.9b04281} or promote \cite{doi:10.1021/acs.jpcc.0c04913} 2D ferromagnetism. VSe$_2$ is a paradigmatic example of this as, despite some early claims \cite{Bonilla2018}, it is now clear both experimentally and theoretically that the CDW order quenches the emergence of itinerant ferromagnetism \cite{PhysRevLett.121.196402, PhysRevB.104.125430,doi:10.1021/acs.jpcc.9b04281,doi:10.1021/acs.nanolett.0c04409, doi:10.1021/acs.nanolett.8b01764,doi:10.1021/acs.nanolett.8b01649, PhysRevB.101.035404,doi:10.1021/acs.jpcc.9b08868}. 

In its bulk form, VSe$_2$ develops a commensurate  $4 \times 4 \times 3$ CDW phase below 110 K \cite{Eaglesham_1986}. The CDW phase opens pseudogaps at the Fermi level impeding the emergence of ferromagnetism \cite{doi:10.1021/acs.jpcc.9b08868}. 
Inelastic x-ray scattering experiments and non-perturbative anharmonic phonon calculations have proven that the CDW transition is driven by the collapse of a low-energy acoustic mode and that the electron-phonon coupling is the origin of the instability \cite{Diego2021}, as suggested as well by other quantitative models \cite{Henke2020}. These anharmonic calculations have shown that van der Waals interactions are essential to melt the CDW and obtain a charge density wave temperature (T$_{CDW}$) in agreement with experiments. This suggests that the CDW in the monolayer may also be characterized by similar phonon softening effects, but with limited influence of van der Waals interactions. 

The main problem in the monolayer of VSe$_2$ is that the CDW is not fully understood yet, as unrelated experiments have reported  distinct CDW orders with different transition temperatures.
A non-monotonic evolution of T$_{CDW}$ as a function of the number of layers has been reported in Ref. \cite{https://doi.org/10.1002/anie.201304337,doi:10.1063/1.4893027, P_sztor_2017}, but retaining an in-plane $4 \times 4$  modulation. A metastable phase with modulation  $4 \times \sqrt{3}$ has also been identified for the few-layer case \cite{PhysRevMaterials.1.024005}.
In the purely 2D limit different CDW orders with non-equivalent modulations have been found. A $4 \times 4$ order was observed in VSe$_2$ films grown on bilayer graphene on top of SiC and on highly oriented pyrolytic graphite (HOPG), with a T$_{CDW}$ of $\sim 140 \pm 5$ K and a lattice parameter of $a = 3.31 \pm 0.05$ \AA \cite{doi:10.1021/acs.nanolett.8b01649}. On the contrary,  a $\sqrt{3} \times \sqrt{7}$ modulation has been observed in VSe$_2$ samples grown on several substrates by molecular beam epitaxy by different groups,  with a consistent T$_{CDW} = 220$ K \cite{PhysRevLett.121.196402,doi:10.1021/acs.jpcc.9b04281}. Some other orders have also been reported: a combination of $2 \times \sqrt{3}$ and $\sqrt{3}$ $\times$ $\sqrt{7}$ with a T$_{CDW}\sim$ 135 K  \cite{doi:10.1021/acs.nanolett.8b01764,doi:10.1021/acs.nanolett.0c04409}, and a $4 \times 1$ modulation with T$_{CDW}\sim$ 350 K \cite{doi:10.1021/acs.nanolett.8b01764,Duvjir_2021}. These experimental contradictions point to the presence of different competing CDW orders, which can lead to different low-temperature phases depending on the substrate  \cite{doi:10.1021/acs.nanolett.8b01764, Duvjir_2021}. 

By calculating the harmonic phonons of the VSe$_2$ monolayer within density functional theory (DFT), theoretical studies have also described the competition of different CDW orders and how strain can influence the ground state \cite{PhysRevB.101.235405}. Harmonic phonon calculations, however, cannot explain that above T$_{CDW}$ the 1T phase is the ground state. In the presence of competing orders, only calculations considering anharmonicity can disentangle what is the CDW order and the transition temperature, as it has already been shown in different transition metal dichalcogenides (TMDs) \cite{doi:10.1021/acs.nanolett.9b00504, PhysRevLett.125.106101,Diego2021,Zhou2020,Sky_Zhou_2020}. Therefore, in order to unveil the intrinsic CDW orders of monolayer VSe$_2$ and how they are affected by external fields, a DFT study including anharmonicty is required.

In this work, we present a theoretical analysis of the CDW orders arising in monolayer VSe$_2$ using non-perturbative anharmonic phonon calculations based on the stochastic self-consistent harmonic approximation (SSCHA) \cite{PhysRevB.89.064302,PhysRevB.96.014111,PhysRevB.98.024106,Monacelli2021}. 
This formalism has been crucial to understand and characterize the CDWs in several TMDs \cite{doi:10.1021/acs.nanolett.9b00504, PhysRevLett.125.106101,Diego2021,Zhou2020,Sky_Zhou_2020} as it overcomes the limitations of the harmonic analysis, allowing to determine the dependence of the CDW order as a function of temperature. 
We demonstrate the emergence and competition of two intrinsic CDW orders in monolayer VSe$_2$ as  
for a lattice parameter of $a=3.35$ \AA$ $ a  $\sqrt{3}$ $\times$ $\sqrt{7}$ order dominates with T$_{CDW}=$ 217 K, 
while for slightly smaller $a=3.30$ \AA$ $ the $4$ $\times$ $4$ order prevails with T$_{CDW}=$ 223 K. 
Moreover, the non-perturbative anharmonic procedure allows us to demonstrate that the CDW  can be suppressed by the inclusion of Lennard-Jones energy terms, which might appear naturally or may be artificially induced by the interplay between the monolayer and a particular substrate.  

\begin{figure}[h!]
  \centering
  \includegraphics[width=\columnwidth]
        {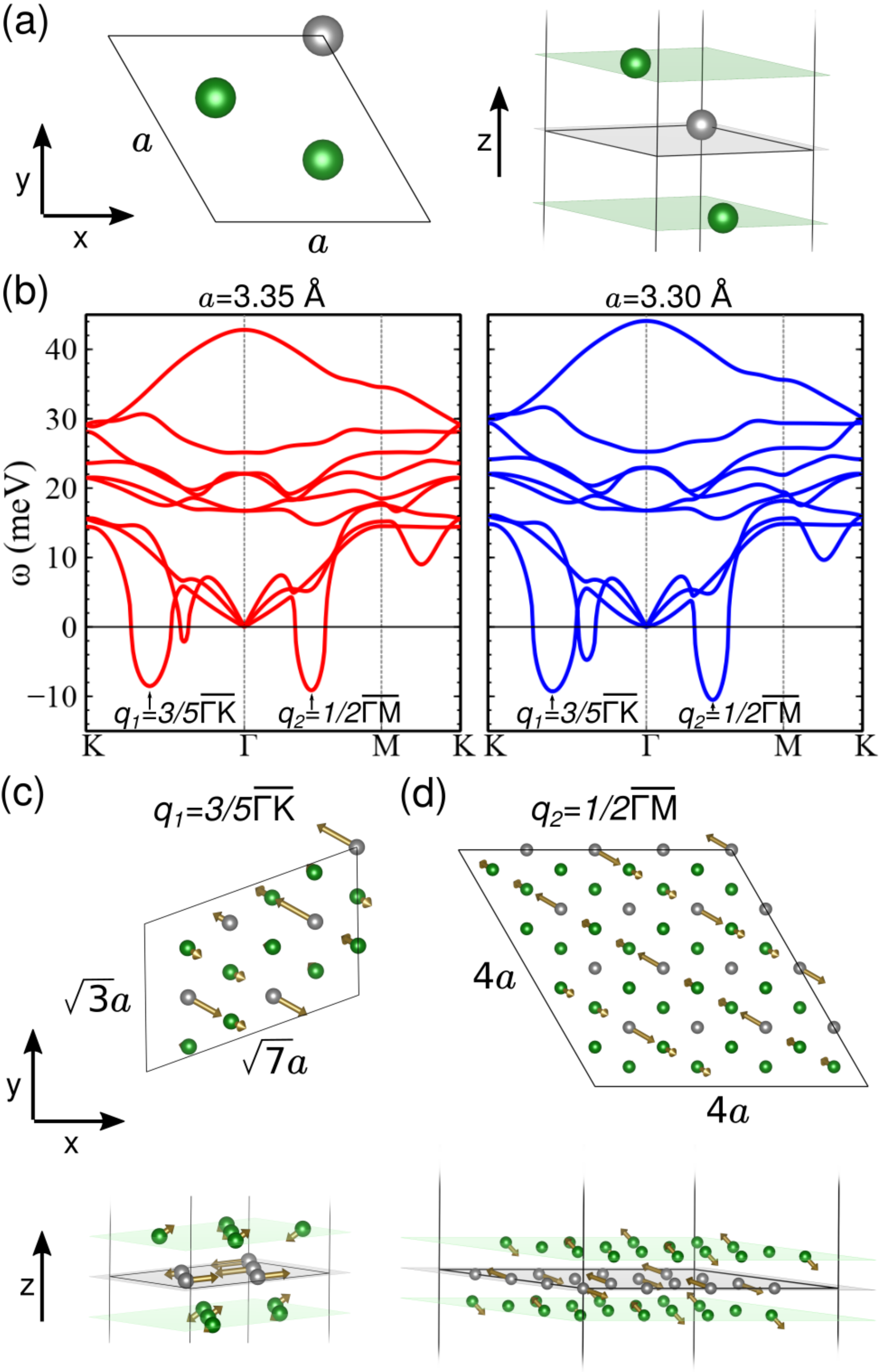}
     \caption{ (a) Normal state structure of monolayer VSe$_2$ with lattice parameter $a$. V (Se) atoms are depicted in gray (green).
     (b) Harmonic phonon band structures of monolayer VSe$_2$ as a function of the lattice parameter, left (right) panel for $a=3.35$ \AA$ $ ($a=3.30$ \AA$ $). 
     Two dominant instabilities at $q_1=3/5\overline{\Gamma \text{K}}$ and $q_2=1/2\overline{\Gamma \text{M}}$ can be identified.  
     (c,d) Intrinsic CDW orders with $\sqrt{3}$ $\times$ $\sqrt{7}$ and $4$ $\times$ $4$ modulations associated to the instabilities at $q_1$ and $q_2$ can be identified in the harmonic phonon band structures. The displacement vectors associated to each CDW order are plotted as brown arrows.  Planes perpendicular to the z-direction for V (Se) were plotted in gray (green) for a better characterization of the displacement vectors.}
     \label{Fig:harm_phonons}
\end{figure}

\begin{figure*}
  \centering
  \includegraphics[width=\textwidth]
        {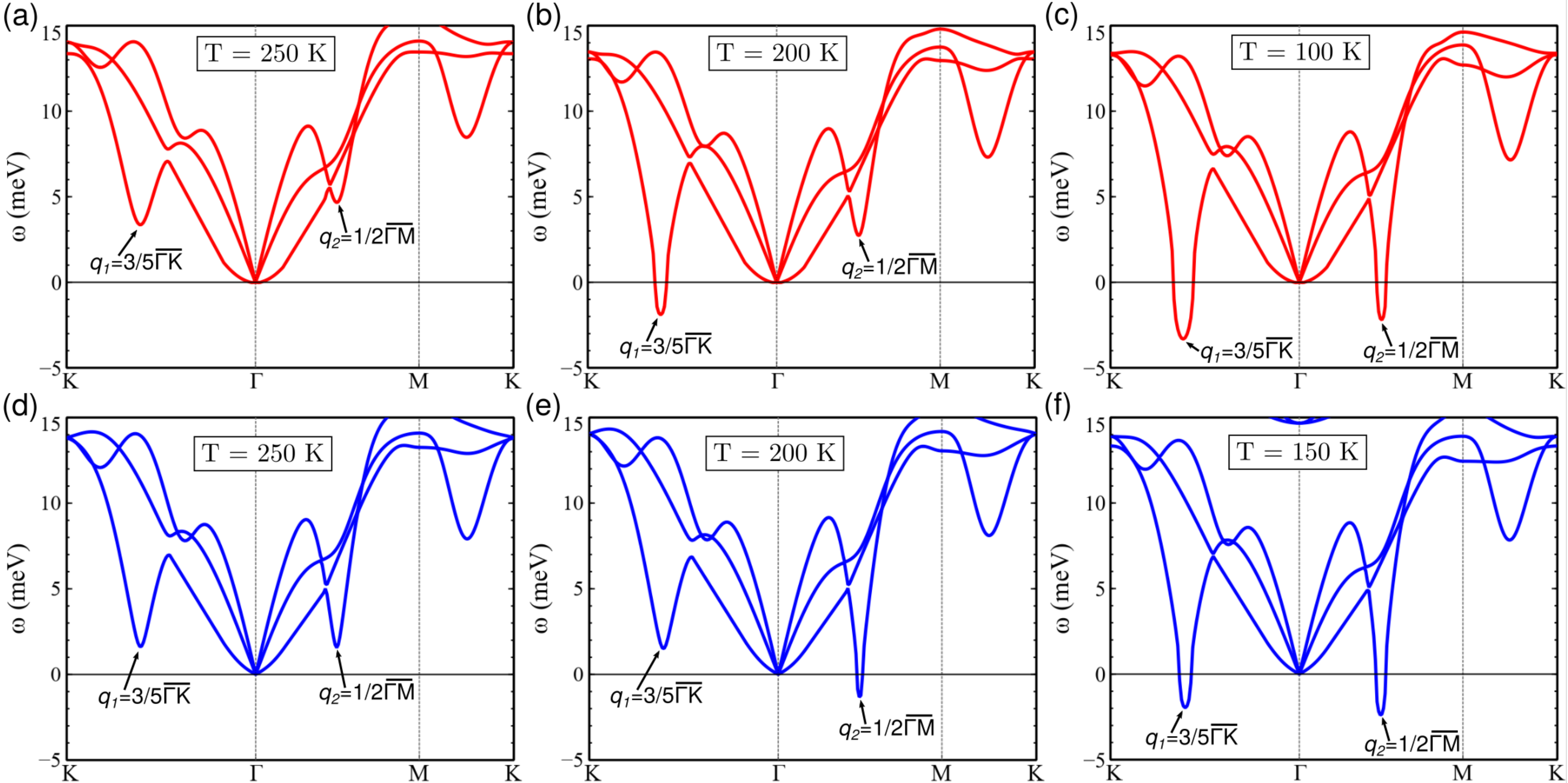}
     \caption{Non-perturbative anharmonic calculations of the NS of monolayer VSe$_2$. (a,b,c) For lattice parameter $a=3.35$ \AA$ $ and temperatures 250, 200, 100 K respectively. 
     (d,e,f) For lattice parameter $a=3.30$ \AA$ $ and temperatures 250, 200, 150 K. 
     The phonon melting occurs first at the $q_1$ ($q_2$) point for $a=3.35$ \AA$ $ ($a=3.30$ \AA$ $) as shown in panels (b,e).}
     \label{Fig:anharm_phonons_T}
\end{figure*}

\section{CDW orders and their transition temperatures}

We start our analysis performing harmonic phonon calculations on monolayer VSe$_2$. The normal state (NS) unit cell is shown in Fig. \ref{Fig:harm_phonons}(a). 
The value of the experimentally reported lattice parameter $a = 3.31 \pm 0.05$ \AA\ is  in rather good agreement with the theoretical one of 3.35 \AA\ obtained at the Perdew-Burke-Ernzerhof \cite{PhysRevLett.77.3865} level without considering the zero point motion \cite{PhysRevLett.121.196402,doi:10.1021/acs.nanolett.8b01649}. Therefore, in this study we perform calculations for two lattice parameters $a=3.35$ \AA$ $ and $a=3.30$ \AA$ $, which provide a good representation of the experimental range.  Density functional perturbation theory (DFPT) \cite{RevModPhys.73.515} is used to compute the harmonic phonon band structure for both lattice parameters \footnote{See the Supplemental material for a detailed description of the calculations.}.
Both harmonic phonon bands (see Fig. \ref{Fig:harm_phonons}(b)) show two dominant instabilities at $q_1=3/5\overline{\Gamma \text{K}}$ and $q_2=1/2\overline{\Gamma  \text{M}}$. 
These are associated with the two intrinsic CDW orders of monolayer VSe$_2$, with modulations shown in Figs. \ref{Fig:harm_phonons}(c) and \ref{Fig:harm_phonons}(d), that lower its Born-Oppenheimer energy.  
The instability at $q_1$ is associated with a  $\sqrt{3} \times \sqrt{7}$ supercell, while the one at $q_2$ leads to a $4 \times 4$ modulation. Both softened phonon modes have out-of-plane component in the displacement vectors, as it is also the case of the CDW instability in the bulk form of this compound.
In spite of providing the two intrinsic CDW orders, harmonic calculations do not suffice to predict neither which of these CDW orders is the dominant one, 
nor the associated transition temperature for each lattice parameter. In fact, the small change in the lattice parameter does not impact the weight of the instabilities. 
Our non-perturbative anharmonic calculations based on a free energy formalism within the SSCHA can give the answer to these questions \footnote{See the Supplemental material for a detailed description of the SSCHA method and the technical aspects of these calculations.}.

Figure \ref{Fig:anharm_phonons_T} shows the temperature evolution of the phonon band structure obtained with the SSCHA method for the two lattice parameters $a=3.35$ \AA$ $ (in red in the top panels of Fig. \ref{Fig:anharm_phonons_T}) and $a=3.30$ \AA$ $ (in blue in the lower panels of Fig. \ref{Fig:anharm_phonons_T}).
At high enough temperature (T=250 K) the 1T NS phase (Fig. \ref{Fig:harm_phonons}(a)) is dynamically stable for both lattice parameters as shown in Figs. \ref{Fig:anharm_phonons_T}(a) and \ref{Fig:anharm_phonons_T}(d), remarking that anharmonicity melts the CDW phase as it happens in other TMDs \cite{doi:10.1021/acs.nanolett.9b00504, PhysRevLett.125.106101,Diego2021,Zhou2020,Sky_Zhou_2020}. 
By decreasing the temperature, the phonon modes associated with the CDW instabilities at $q_1$ and $q_2$ soften. In particular, for  $a=3.35$ \AA$ $ at 200 K (Fig. \ref{Fig:anharm_phonons_T}(b)) we can observe that the mode at $q_1=3/5\overline{\Gamma \text{K}}$ becomes unstable, even if the one at $q_2$ remains stable. This result indicates that for this lattice parameter the $\sqrt{3} \times \sqrt{7}$ CDW order dominates. However, for  $a=3.30$ \AA$ $ at 200 K (Fig. \ref{Fig:anharm_phonons_T}(e)) the phonon mode  at $q_2=1/2\overline{\Gamma  \text{M}}$ is unstable, but not at $q_1$. Therefore, for the smaller lattice parameter, the $4 \times 4$ CDW order is the dominant one. At low enough temperatures (Figs. \ref{Fig:anharm_phonons_T}(c) and \ref{Fig:anharm_phonons_T}(f)) both $q$-vectors show unstable modes. However, note that this situation is not indicating that at low temperatures both CDW orders coexist, albeit it is a clear signature that the anharmonic free energy landscape becomes more complex. Once one of the CDW orders gets stable when decreasing the temperature, the system collapses to it, and the analysis in terms of the anharmonic phonons of the NS phase is no longer useful to describe the evolution of each of the CDW phases at low temperatures. Nevertheless, the anharmonic phonons at low temperature shown in Figs. \ref{Fig:anharm_phonons_T}(c) and \ref{Fig:anharm_phonons_T}(f) confirm that both $q_1$ and $q_2$ are the intrinsic CDW orders of VSe$_2$ that can be accessed through a transition from the NS phase.

\begin{figure}[h!]
  \centering
  \includegraphics[width=\columnwidth]
        {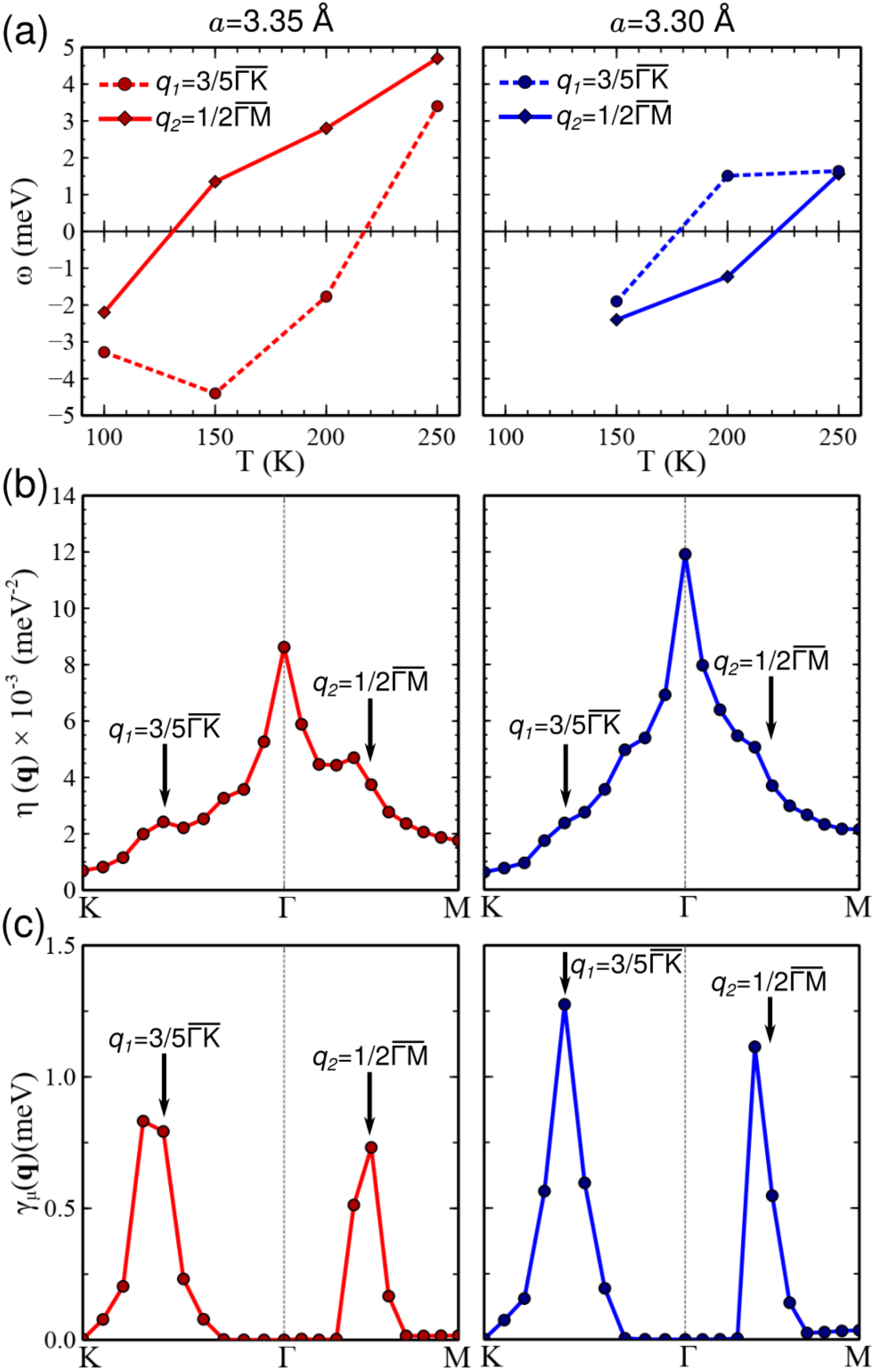}
     \caption{(a) Temperature evolution of the frequencies of the softened modes at $q_1$ and $q_2$ as obtained in the SSCHA calculation. 
     (b) Nesting function $\eta(q)$. It does not clearly peak at $q_1$ and $q_2$. 
     (c) Phonon linewidth $\gamma_{\mu}(\textbf{q})$
given by the electron-phonon interaction. Sizable peaks appear at the CDW $q$-vectors. All the results for $a=3.35$ \AA$ $ ($a=3.30$ \AA$ $) are shown in red (blue) in the left (right) panels.}
     \label{Fig:cdw_T_elph_nest}
\end{figure}

To analyze in more detail the competition between the two CDW orders as a function of the lattice parameter, Fig. \ref{Fig:cdw_T_elph_nest}(a) shows the temperature evolution of the frequency of the phonon mode that softens at  $q_1$ and $q_2$. 
For the larger lattice parameter, $a=3.35$ \AA$ $, the frequency at $q_1$ becomes negative (imaginary) at higher temperature than at $q_2$, and hence the $\sqrt{3}$ $\times$ $\sqrt{7}$ CDW order is the dominant (left panel in Fig. \ref{Fig:cdw_T_elph_nest}(a) in red). The opposite behavior is observed for the small lattice parameter $a=3.30$ \AA$ $. The frequency at $q_2$ becomes negative at higher temperature that at $q_1$, and hence the $4$ $\times$ $4$ CDW order is dominant (right panel in Fig. \ref{Fig:cdw_T_elph_nest}(a) in blue). From Fig. \ref{Fig:cdw_T_elph_nest}(a) we can obtain the transition temperature for each lattice parameter: for $a=3.35$ \AA$ $ the  $\sqrt{3} \times \sqrt{7}$ order emerges at T$_{CDW}=$ 217 K, while for  $a=3.30$ \AA$ $ the $4$ $\times$ $4$ order arises at T$_{CDW}=$ 223 K. Importantly, our anharmonic calculations including the zero point energy \cite{PhysRevB.98.024106} predict an associated in-plane pressure of 0.7 GPa for $a=3.35$ \AA\ and 1.3 GPa for $a=3.30$ \AA$ $. Considering that its in-plane pressure is lower, these results point out that the intrinsic CDW order in monolayer VSe$_2$ is $\sqrt{3} \times \sqrt{7}$ with a T$_{CDW}=$ 217 K, which is in perfect agreement with the experiments on Refs. \cite{PhysRevLett.121.196402,doi:10.1021/acs.jpcc.9b04281}, 
and that the $4$ $\times$ $4$ order, which is the in-plane projection of the bulk $4$ $\times$ $4$ $\times$ $3$ CDW order, appears only under strain. 
Our results provide an explanation for the different CDW orders observed for small variations ($\sim$1.5\%) of the lattice parameter \cite{PhysRevLett.121.196402,doi:10.1021/acs.nanolett.8b01649}. Note that, eventually, other modulations could appear in monolayer VSe$_2$ as experimentally reported\cite{doi:10.1021/acs.nanolett.8b01764,doi:10.1021/acs.nanolett.0c04409,Duvjir_2021}. However, our results show that the $\sqrt{3} \times \sqrt{7}$ and $4 \times 4$ modulations are the intrinsic CDW orders, and point out that those different modulations are a consequence of the interplay between the highly dynamically-unstable NS of VSe$_2$ monolayer and the particular substrate. 

Having established the competition between the two intrinsic CDW orders of monolayer VSe$_2$ as a function of the lattice parameter, we study now the origin of these CDW orders. In order to do so, we use DFPT to compute both the nesting function $\eta(\textbf{q})$ (Fig.  \ref{Fig:cdw_T_elph_nest}(b)), which is given by 
\begin{equation}
    \eta(\textbf{q})=\frac{1}{N}\sum_{nn'}\sum_{\textbf{k}}^{1BZ} \delta(\epsilon_{n'\textbf{k}+\textbf{q}}-\epsilon_{F})\delta(\epsilon_{n\textbf{k}}-\epsilon_{F}) \;, \label{eq:nesting}
\end{equation}
and the phonon linewidth associated to  the electron-phonon interaction (see Fig. \ref{Fig:cdw_T_elph_nest}(c)),
\begin{equation} \small
\gamma_{\mu}(\textbf{q})=\frac{2\pi w_{\mu}(\textbf{q})}{N}\sum_{nn'}\sum_{\textbf{k}}^{1BZ}|g^{\mu}_{n'\textbf{k}+\textbf{q},n\textbf{k}}|^2 \delta(\epsilon_{n'\textbf{k}+\textbf{q}}-\epsilon_{F})\delta(\epsilon_{n\textbf{k}}-\epsilon_{F}) \;.
\label{eq:elph}
\end{equation}
In Eqs. \eqref{eq:nesting} and \eqref{eq:elph} $\epsilon_{n\textbf{k}}$ is the energy of band $n$ with wave number $\textbf{k}$, $\epsilon_{F}$ the Fermi energy, and $N$ the number of $\textbf{k}$ points in the sum over the first Brillouin zone (1BZ). The equation for $\gamma_{\mu}(\textbf{q})$ is very similar to the nesting function, but the value is weighted by the mode $\mu$ and momentum $\textbf{q}$ dependent  electron-phonon matrix elements $g^{\mu}_{n'\textbf{k}+\textbf{q},n\textbf{k}}$. It is worth noting that the electron-phonon linewidth is independent of the phonon frequency $w_{\mu}(\textbf{q})$ as the electron-phonon matrix elements scale as  $w_{\mu}(\textbf{q})^{-1/2}$. The nesting function peaks for $\mathbf{q}$ points that connect nested regions of the Fermi surface, and, thus, reveals if the instability emerges from a purely electronic instability. In principle $\gamma_{\mu}(\textbf{q})$ also peaks at these $\mathbf{q}$ points, but it is affected by momentum dependent electron-phonon matrix elements.
Therefore, the direct comparison between the two allows us to establish which is the main driving force of the CDW orders in VSe$_2$ monolayer. 

We can see in Fig.  \ref{Fig:cdw_T_elph_nest}(b) that the nesting function does not show any peak at the CDW vectors for both lattice parameters, despite the existence of small shoulders near $q_1$ and $q_2$. Therefore, the CDW vectors do not coincide with a nested region of the Fermi surface. In a different way, Fig. \ref{Fig:cdw_T_elph_nest}(c) shows that the phonon linewidth coming from the electron-phonon interaction abruptly peaks at both $q_1$ and $q_2$ for both lattice parameters, meaning that in all cases the enhancement comes from the mode and momentum dependence of the electron-phonon matrix elements. In conclusion, the 2 intrinsic CDW orders developed by monoloyer VSe$_2$ are driven by the electron-phonon coupling, without the need for any nesting mechanism. Precisely, this result is in agreement with theoretical analyses that claim that, for 2D systems, purely electronic interactions do not suffice to produce a CDW order, and electron-phonon interactions are the main driving force in this kind of transition \cite{PhysRevB.77.165135}. Besides, the electron-phonon interaction also plays a key role in the CDW transition in 3D systems, as in bulk 1\textit{T}-VSe$_2$, in which case the presence of nesting is symbolic \cite{Diego2021}.

\section{van der Waals interactions and the suppression of the CDW instabilities}

The analysis about the stability of the different CDW orders as a function of strain was performed with a non-local van der Waals density exchange-correlation functional \cite{PhysRevB.76.125112}. The reason for this is that this functional allows to properly describe both bulk and monolayer limits of VSe$_2$, oppositely to the widely used GGA-PBE functional \footnote{See the Supplemental material for a more detailed description of the election of the exchange correlation functional to study the CDW orders of VSe$_2$.}.
In the latter case, a huge overestimation of T$_{CDW}$ occurs in bulk due to the lack of van der Waals interactions that cause a melting of the CDW phase \cite{Diego2021}.
Motivated by these results in the bulk, we explore here the effect that external van der Waals interactions may cause in the CDW orders of monolayer VSe$_2$. These van der Waals interactions might naturally appear by proximity effect between the analyzed monolayer and other layers, such as the substrate or in van der Waals heterostructures.

\begin{figure}[h!]
  \centering
  \includegraphics[width=\columnwidth]
        {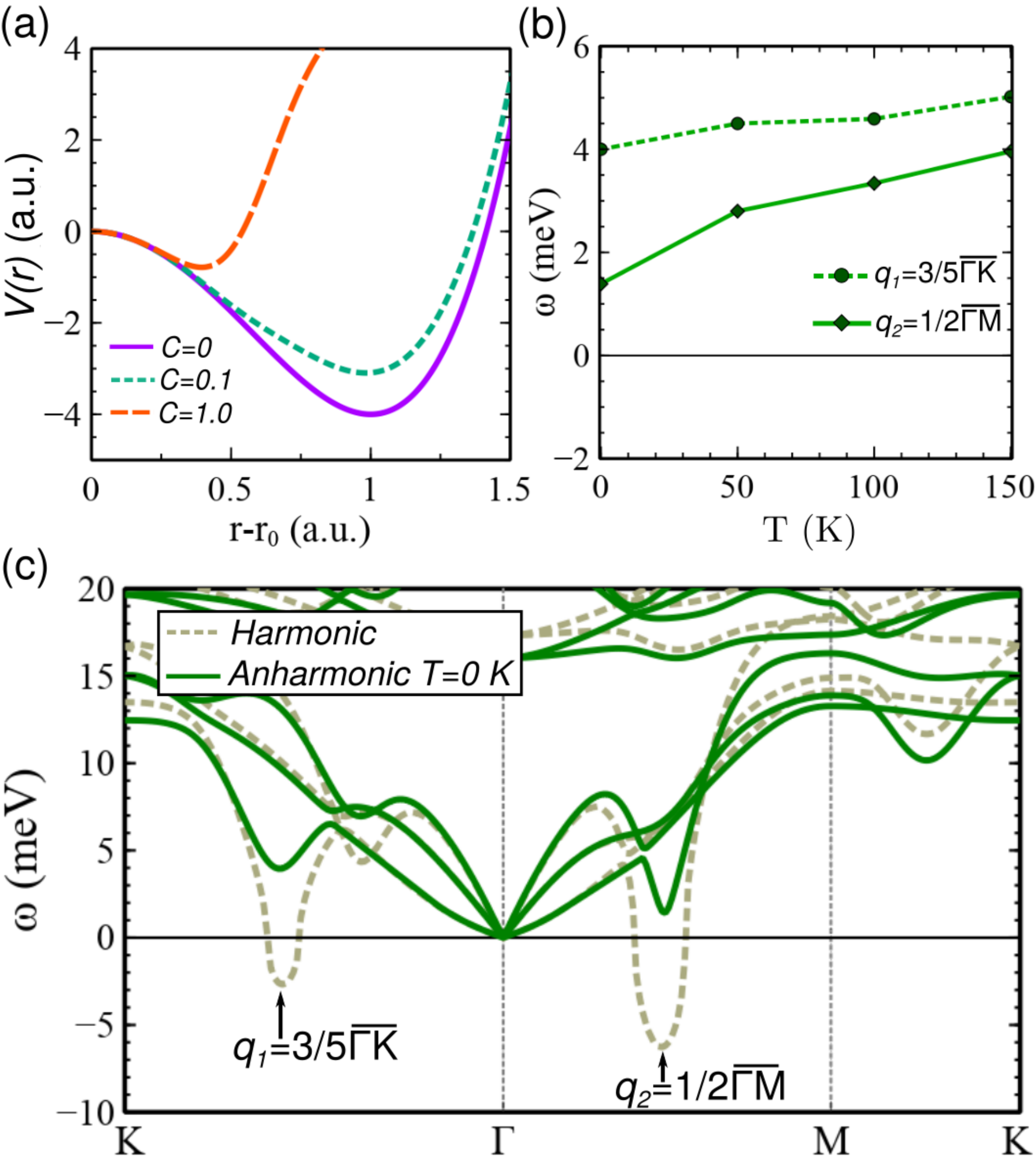}
     \caption{(a) $V(r)$ potential as a function of the atomic position $r$ described by Eqs. \eqref{eq:pes} and \eqref{eq:lj} in arbitrary units (a.u.) for different $C$ values. Particular values of $A=-8$, $B=-A/2$, and $r_c=0.1$ are considered.
     (b) Temperature evolution of the soften modes' frequencies at $q_1$ and $q_2$ when anharmonicity and Lennard-Jones interactions are included. 
     (c) Harmonic and anharmonic phonon band structures at T=0 K for monolayer VSe$_2$ including Lennard-Jones contributions.}
     \label{Fig:lj_interactions}
\end{figure}

In order to illustrate the effect of external van der Waals forces on the CDW, we make use of a simple one dimensional double-well fourth order potential: 
\begin{equation}\label{eq:pes}
    V(r)=A\left( r-r_0\right)^2+B\left( r-r_0\right)^4,
\end{equation}
where $A$ and $B$ are the coefficients of the different powers, $r$ the position of the atoms, and $r_0$ is the equilibrium atomic position in the high temperature phase. The CDW occurs when the free energy calculated with this potential is lower at $r-r_0\ne 0$ than at $r-r_0 = 0$. Obviously the lower and wider the minimum of the well, the more probable to find a broken-symmetry CDW order. We can now add to the potential $V(r)$ a $E_{LJ}(r)$ Lennard-Jones energetic contribution to mimic the role played by external van der Waals interactions: 
\begin{equation}\label{eq:lj}
    E_{LJ}(r)=\frac{C}{r_c+\left( r-r_0\right)^6}+D,
\end{equation}
where $C$ is the coefficient that controls the strength of the Lennard-Jones interactions, $r_c$ is a cut-off radius that prevents a divergence at $r=r_0$, and $D=C/r_c$ is simply a constant  that fixes the potential to be equal to 0 at $r_0$. The effect of  the Lennard-Jones interactions on $V(r)$ can be seen in Fig. \ref{Fig:lj_interactions}(a) for different values of $C$. We can observe that an increase in the strength of the Lennard-Jones interactions makes the potential shallower. Therefore, Lennard-Jones interactions tend to quench the low-temperature CDW phase and promote the high-temperature symmetric phase.

We can confirm this simple picture in the particular case of monolayer VSe$_2$ by including an energy term like the one shown in Eq. \eqref{eq:lj} through the Grimme's semiempirical approach in our SSCHA calculations on top of the PBE functional \cite{doi:10.1002/jcc.20495}. Figure \ref{Fig:lj_interactions}(c) shows the harmonic phonon band structure including energy contributions from Lennard-Jones interactions \footnote{These calculations are  for $a=3.35$ \AA$ $, but qualitatively the results hold for any other lattice parameter. The strength of the Lennard-Jones interactions was set to the one considered by default by the Grimme's semiempirical correction implemented in the {\sc Quantum ESPRESSO} package \cite{0953-8984-21-39-395502,0953-8984-29-46-465901}}. 
The instability at  $q_1$ and $q_2$ is reduced compared to the harmonic bands without the Lennard-Jones contribution shown \footnote{See the Supplemental material for a better visualization of this effect at the harmonic level.}. In this situation, anharmonic effects are able to suppress both CDW orders by stabilizing the softened phonons even at 0 K, as shown in Fig. \ref{Fig:lj_interactions}(c). 
This effect can be also analyzed in Fig. \ref{Fig:lj_interactions}(b), where the evolution of the frequencies of the softened modes at $q_1$ and $q_2$ as a function of temperature is shown. The frequencies remain stable at any temperature. Therefore, this demonstrates that the combination of Lennard-Jones interactions and anharmonicity can destroy the CDW orders, stabilizing the NS phase at low temperatures. Note that here we have considered a strength of the Lennard-Jones interactions that quenches both CDW orders. However, this strength could be modulated by the parameter $C$ in Eq. \eqref{eq:lj}, providing simply a decrease of  T$_{CDW}$, but not a suppresion of the CDW order as reported for bulk \cite{Diego2021}. Therefore, the effect of Lennard-Jones interactions might be related with the different transition temperatures and CDW orders reported in experiments where VSe$_2$ is grown on different substrates \cite{doi:10.1021/acs.nanolett.8b01649,PhysRevLett.121.196402,doi:10.1021/acs.nanolett.8b01764,doi:10.1021/acs.nanolett.0c04409}, and also may explain the enhancement of the CDW order in the 2D limit \cite{P_sztor_2017}, in which this kind of interactions decrease.



Finally, note that monolayer VSe$_2$ has attracted great attention due to its proximity to an itinerant ferromagnetic state, which is suppressed by the presence of CDW orders \cite{doi:10.1021/acs.jpcc.9b08868,doi:10.1021/acs.jpcc.9b04281}. Our analysis suggests that a ferromagnetic state in monolayer VSe$_2$ may be possible if its intrinsic CDW orders are suppressed by external Lennard-Jones interactions. This tuning could be implemented by substrate engineering or by artificial design of van der Waals heterostructures. In particular, combining compounds that display CDW orders with ferroelectric materials, which  provide strong dipolar interactions, might allow an electric control of CDW phases and the emergence of other competing orders. This mechanism to tune or destroy CDW orders is generic and could be extended to other similar systems, offering a novel platform to engineering new functional materials.


\section{Conclusions}

In conclusion, in this work we  analyze the CDW orders of monolayer VSe$_2$ using non-perturbative anharmonic phonon calculations that allow to determine 
the CDW orders of this system and their corresponding transition temperatures. We  demonstrate the competition between two intrinsic CDW orders as a function of the lattice parameter. Variations of 1.5\% in the lattice parameter are enough to drive the system from the $\sqrt{3} \times \sqrt{7}$ to the $4 \times 4$ order. Transition temperatures on the order of 220 K have been found for both CDW orders, in very good agreement with experiments. 
The role played by external Lennard-Jones interactions in suppressing CDWs described here paves the way to tune CDW orders occurring in van der Waals materials, thus promoting competing orders that might arise in these systems.


\section*{Acknowledgements}

We acknowledge the computational resources provided by the CESGA and the Aalto Science-IT project. 
A.O.F. thanks the financial support received through the Academy of Finland Project No. 349696. J.D. is
also thankful to the Department of Education of the Basque Government for a predoctoral fellowship (Grant No. PRE-2020-1-0220). We thank the Ministry of Science and Education of Spain for financial support through the projects PGC2018-101334-A-C22, GC2018-101334-B-C21, PID2021-122609NB-C22.

\end{document}


\title{Supplemental material: Anharmonicity reveals the tunability of the charge density wave orders in monolayer VSe$_2$} 

\author{Adolfo O. Fumega}
\email{adolfo.oterofumega@aalto.fi}
\affiliation{Department of Applied Physics, Aalto University, 02150 Espoo, Finland}

\author{Josu Diego}
\affiliation{Fisika Aplikatua Saila, Gipuzkoako Ingeniaritza Eskola, University of the Basque Country (UPV/EHU), San Sebastián, Spain}
\affiliation{Centro de Física de Materiales (CSIC-UPV/EHU), San Sebastián, Spain}

\author{V. Pardo}
\affiliation{Departamento de Física Aplicada, Universidade de Santiago de Compostela, Santiago de Compostela, Spain}
\affiliation{Instituto de Materiais iMATUS, Universidade de Santiago de Compostela, Santiago de Compostela, Spain}

\author{S. Blanco-Canosa}
\affiliation{Donostia International Physics Center (DIPC), San Sebastián, Spain}
\affiliation{IKERBASQUE, Basque Foundation for Science, 48013 Bilbao, Spain}

\author{Ion Errea}
\email{ion.errea@ehu.eus}
\affiliation{Fisika Aplikatua Saila, Gipuzkoako Ingeniaritza Eskola, University of the Basque Country (UPV/EHU), San Sebastián, Spain}
\affiliation{Centro de Física de Materiales (CSIC-UPV/EHU), San Sebastián, Spain}
\affiliation{Donostia International Physics Center (DIPC), San Sebastián, Spain}

\maketitle

\section{Computational methods}

\subsection{Density Functional Perturbation Theory (DFPT) calculations }

Harmonic phonon frequencies and electron-phonon matrix elements were calculated within density functional perturbation theory (DFPT) \cite{RevModPhys.73.515} as implemented in the {\sc Quantum ESPRESSO} package \cite{0953-8984-21-39-395502,0953-8984-29-46-465901}. The force calculations needed for DFPT calculations were performed making use of a non-local van der Waals exchange-correlation functional \cite{HK,PhysRevB.76.125112}. We used an ultrasoft pseudopotential that includes 4$s^2$ 3$d^3$ valence electrons for V and a norm-conserving one with 4$s^2$ 4$p^4$ electrons in the valence for Se. We used a plane-wave energy cutoff of 50 Ry for the wavefunctions and 550 Ry for the charge density. The Brillouin zone integrals were performed in a 32$\times$32$\times$1 k-point grid with a Methfessel-Paxton smearing \cite{MP} of 0.01 Ry. Harmonic phonon calculations were carried out in a 8$\times$8$\times$1 q-point grid. The nesting function and the electron-phonon linewidth were calculated using a 48$\times$48$\times$1 k-point grid and a Gaussian broadening of 0.003 Ry for the Dirac deltas. 

\subsection{The stochastic self-consistent harmonic approximation (SSCHA)}

The stochastic self-consistent harmonic approximation  (SSCHA) \cite{PhysRevB.89.064302,PhysRevB.96.014111,PhysRevB.98.024106, Monacelli2021} is a quantum variational method on the free energy fully accounting for anharmonic effects at any temperature. The variational minimization is carried out with respect to a trial harmonic density matrix $\rho_{\mathcal{H}}$ that contains two groups of parameters: the force-constants $\boldsymbol{\Phi}$ and the \textit{centroid} positions $\boldsymbol{\mathcal{R}}$. The centroid positions at the free energy minimum $\boldsymbol{\mathcal{R}}_{eq}$ are the average ionic equilibrium positions fully accounting for quantum, thermal and anharmonic effects. The temperature-dependent anharmonic phonon spectra at the static level are obtained from the diagonalization of a free energy Hessian based dynamical matrix, ${D}_{ab}^{(F)}=\frac{1}{\sqrt{M_aM_b}}	\frac{\partial^{2} F}{\partial \mathcal{R}_a \partial \mathcal{R}_b}\Big|_{\mathcal{R}_{eq}}$\cite{PhysRevB.96.014111}. 
Following Landau’s theory \cite{landau}, these phonons are enough to identify second-order structural phase transitions, being the transition critical temperature the one in which a phonon mode goes to null frequency. 

The variational free energy minimization within the SSCHA method was performed by calculating forces on 4$\times$4$\times$1 supercells making use of DFT as implemented in {\sc Quantum ESPRESSO}. We used the same non-local van der Waals exchange-correlation functional, pseudopotentials and parameters described in the previous section, but with a 4$\times$4$\times$1 grid in the unit cell for the Brillouin zone integrals. The theoretical anharmonic phonon spectra shown in the paper were calculated in the static limit of the SSCHA theory; based on the free energy Hessian formalism. The difference between anharmonic and harmonic dynamical matrices was interpolated to a finer grid of size 8$\times$8$\times$1  in order to obtain other anharmonic phonon frequencies in more \textbf{q} points. 

\section{Election of the exchange correlation functional}

Importantly, in the anharmonic analyses carried out in this manuscript with the SSCHA methodology, we have considered the a non-local van der Waals exchange-correlation functional \cite{PhysRevB.76.125112}. In this section we justify such a choice.

\begin{figure}[h!]
  \centering
  \includegraphics[width=\columnwidth]
        {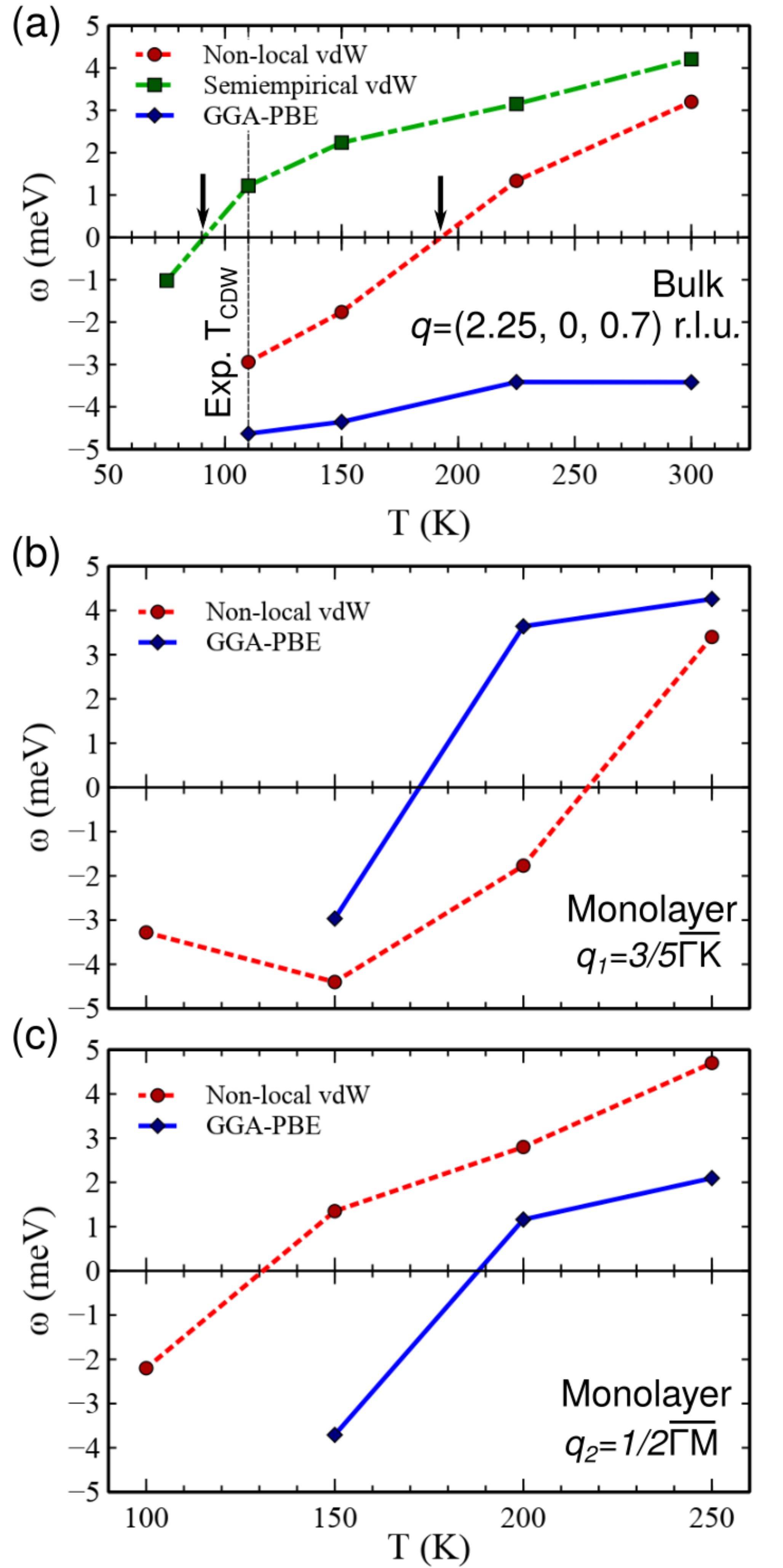}
     \caption{Temperature evolution of the frequencies of the softened modes associated to the CDW orders for different exchange correlation functionals: (a) Evolution for bulk VSe$_2$  at $q=$(2.25,0.0,0.7), data for the semiempirical van der Waals and the GGA-PBE functional taken from Ref. \cite{Diego2021}. The predicted transition temperatures for each functional are highlighted with an arrow, the experimental  T$_{CDW}$ is also indicated. No transition to the NS is observed below 300K for the GGA-PBE functional. (b,c) Evolution for monolayer VSe$_2$ at $q_1=3/5\overline{\Gamma \text{K}}$ and $q_2=1/2\overline{\Gamma \text{M}}$ associated to the $\sqrt{3}$ $\times$ $\sqrt{7}$ and $4$ $\times$ $4$ modulations respectively considering  $a=3.35$ \AA$ $.}
     \label{Fig:functional_bulk_mono}
\end{figure}

As reported in Ref. \cite{Diego2021}, it is fundamental to include van der Waals interactions to provide a good description of the transition temperature from the NS to the CDW state. Figure \ref{Fig:functional_bulk_mono}a shows a summary of the results for bulk VSe$_2$ obtained in Ref. \cite{Diego2021} using the GGA-PBE exchange-correlation functional \cite{PhysRevB.54.16533} and including the van der Waals interactions in a semiempirical way \cite{doi:10.1002/jcc.20495}. Moreover, we have included our calculations using the non-local van der Waals exchange-correlation functional \cite{PhysRevB.76.125112}. We can observe that the prediction using GGA-PBE is totally wrong since it does not show any transition from the CDW to the NS state at temperatures below 300 K, being  T$_{CDW}=110$ K. The reason for this result is that van der Waals interactions are not taken into account at the PBE level. In contrast, we can observe that for the semiempirical approach and for the non-local van der Waals functional the phase transition is captured below 200 K. In particular, the non-local van der Waals functional  predicts a transition temperature of T$_{CDW}=190$ K, this overestimation might be due to the lack of some non-local effects that are difficult to capture in a DFT exchange-correlation functional. The semiempirical approach is based on Lennard-Jones terms that are added in a non-\emph{ab initio} way. For the default values set in the {\sc Quantum ESPRESSO} package for the strength of this interaction T$_{CDW}=90$ K, thus producing a small underestimation of the transition temperature. Of course in this semiempirical approach the strengh of the van der Waals interactions could be varied to describe better the transition, however note that this would depart from an \emph{ab initio} determination of the T$_{CDW}$.

Figures \ref{Fig:functional_bulk_mono}b and \ref{Fig:functional_bulk_mono}c  show the temperature evolution of the frequencies of the softened modes at $q_1=3/5\overline{\Gamma \text{K}}$ and $q_2=1/2\overline{\Gamma \text{M}}$ associated to the CDW orders found in monolayer VSe$_2$. Results for the GGA-PBE and the non-local exchange correlation functional are shown using an in-plane lattice parameter  $a=3.35$ \AA$ $ in the calculations. We can see in these figures the predicted dominant CDW order and its corresponding T$_{CDW}$. In the monolayer limit, one might expect that van der Waals interactions are going to be highly suppressed and hence it is going to be irrelevant whether considering the PBE or the non-local approximation. However, there are a couple of important reasons to select the non-local van der Waals functional in the monolayer. First, as we have seen, the non-local functional is able to describe both the bulk and the monolayer limit. This allows us to drive conclusions about the trend of the CDW order as a function of the dimensionality. Our calculations using non-local van der Waals predict an enhancement of the CDW order in the monolayer limit of around $\sim$20-30 K, in good agreement with experimental results \cite{P_sztor_2017}. Second, we can observe from Figs. \ref{Fig:functional_bulk_mono}b and \ref{Fig:functional_bulk_mono}c that the PBE functional predicts the 4 $\times$4 CDW order to be the dominant one at $a=3.35$ \AA$ $, while the non-local functional predicts the $\sqrt{3}$ $\times$ $\sqrt{7}$ to be the dominant one. This disagreement between both functionals stems from the fact that the polarization vectors associated to the CDW orders have out-of-plane components (see Figs. 1c and 1d of the main text) that may induce non-local effects in the monolayer.
Taking into account all this, plus the fact that monolayer VSe$_2$ presents a highly dynamically-unstable NS
justifies the choice of the non-local van der Waals functional as the best choice to determine the competition between both CDW orders. Indeed, we found that the predictions of this functional are in agreement with the reported experimental results  \cite{doi:10.1021/acs.nanolett.8b01649,PhysRevLett.121.196402}.
Finally, notice that using the semiempirical approach to drive conclusions about the intrinsic CDW orders and their competition in the monolayer is not an option, since this treatment is not \emph{ab initio} and the intrinsic non-local effects that might arise in the monolayer cannot be anticipated. However, it can be used to study the qualitative effect of adding external van der Waals interaction to the monolayer, for instance by proximity effects of the monolayer with a substrate or in a van der Waals heterostructure, as we have performed in the main text when considering the effect of external Lennard-Jones energy terms entering in the monolayer.

\section{Effect of external Lennard-Jones interactions at the harmonic level}

\begin{figure}[h!]
  \centering
  \includegraphics[width=\columnwidth]
        {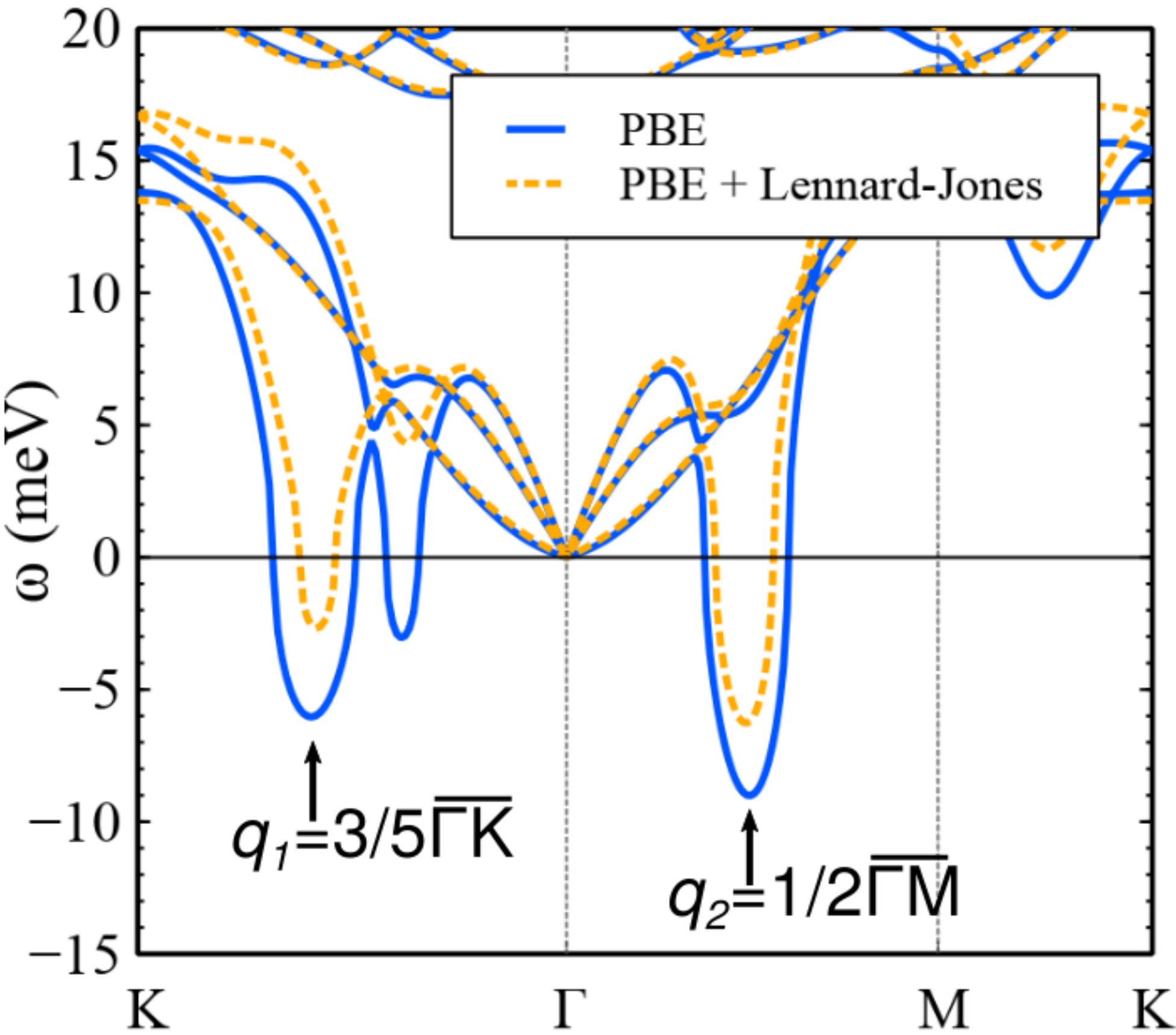}
     \caption{Harmonic phonon calculations for monolayer VSe$_2$. We can observe that including external Lennard-Jones interactions decreases the instability of the unstable modes associated to the CDW orders at $q_1$ and $q_2$.}
     \label{Fig:phonons_harm_pbe_lj}
\end{figure}

The effect of adding external Lennard-Jones interactions in monolayer VSe$_2$ can be seen in Fig. \ref{Fig:phonons_harm_pbe_lj}. We can observe that, at the harmonic level, Lennard-Jones interactions reduce the instabilities associated to the CDW orders. As explained in the main text, this is the expected behavior, since this kind of energy terms lead to the renormalization of the potential energy surface, approaching the system to the high-temperature, symmetric, normal state phase.
 
